\documentclass[aps,prl,twocolumn,showpacs,amssymb,letterpaper]{revtex4}

\usepackage{amsmath} 
\usepackage{epsfig}

\newcommand{\phih}{\widehat{\phi}} 
\newcommand{\bk}{\boldsymbol{k}} 
\newcommand{\bp}{\boldsymbol{p}} 
\newcommand{\bq}{\boldsymbol{q}} 

\begin{document}

\title{Cooperativity Beyond Caging: Generalized Mode Coupling Theory} 

\author{Peter Mayer}
\author{Kunimasa Miyazaki}
\author{David R.\ Reichman}

\affiliation{Department of Chemistry, Columbia University, 3000
Broadway, New York, NY 10027, USA}

\date{\today}

\begin{abstract} 
The validity of mode coupling theory (MCT) is restricted by an 
uncontrolled factorization approximation of density correlations. 
The factorization can be delayed and ultimately avoided, however, 
by explicitly including higher order correlations. We explore this 
approach within a microscopically motivated schematic model. 
Analytic tractability allows us to discuss in great detail the 
impact of factorization at arbitrary order, including the limit of 
avoided factorization. Our results indicate a coherent picture for 
the capabilities as well as limitations of MCT. Moreover, including 
higher order correlations systematically defers the transition and 
ultimately restores ergodicity. Power-law divergence of the 
relaxation time is then replaced by continuous but exponential growth.
\end{abstract} 


\maketitle

The dynamic scaling in dense supercooled liquids and colloidal suspensions 
is a subject surrounded by controversy. Mode coupling theory (MCT) has 
shaped our understanding through detailed and remarkably successful 
predictions \cite{Gotze}. A salient result of MCT is the existence of 
an ideal glass transition. The latter, however, is predicted to occur at 
substantially higher temperatures or lower densities than that observed in 
the laboratory \cite{Kob}. Furthermore it has been suggested that there are 
``activated processes" not accounted for by MCT which restore ergodicity 
and thus round off the ideal transition \cite{instant,emct}. In order to 
remedy such shortcomings one should scrutinize the approximations made 
within MCT. 

A fundamental quantity in MCT is the two-point correlator 
$F_k(t) = \mathcal{N}^{-1} \langle \rho_{-\bk}(0)\rho_{\bk}(t) \rangle$ 
with $\rho_{\bk}$ the Fourier transform of spatial particle density fluctuations. 
For $\mathcal{N}$ particles interacting through a pair potential 
one finds, using standard projection operator techniques 
\cite{Gotze,balucani1994}, 
\begin{equation}
  \ddot{F}_k(t) + \mu_k F_k(t) + (\gamma_{\bk} * \dot{F}_k)(t)=0. 
  \label{equ:Fkt} 
\end{equation}
Here $(f * g)(t) = \int_0^t d\tau f(\tau) g(t-\tau)$ and dots indicate 
time-derivatives. Further $\mu_k = k_B T k^2/m S_k$ where $k_B$ is 
the Bolzmann constant, $T$ the temperature, $m$ the particle mass and 
$S_k = F_k(0)$ the static structure factor. The memory function 
$\gamma_{\bk}(t)$ is related to the auto-correlation of the fluctuating 
force. Within MCT it is assumed \cite{Gotze} that its dominant contribution  
arises from pair-density modes $\rho_{\bq}\rho_{\bk-\bq}$. A 
projection of the fluctuating force onto this pair subspace gives 
$\gamma_{\bk}(t) \approx \sum_{\bq,\bq'} V_{\bq,\bk-\bq}^{*}V_{\bq',\bk-\bq'} 
F_{\bq,\bk-\bq,\bq',\bk-\bq'}(t)$ where $V_{\bp,\bq}$ represents static 
projections of the fluctuating force and $F_{\bp,\bq,\bp',\bq'}(t) = \langle 
\rho_{-\bp}(0) \rho_{-\bq}(0) \rho_{\bp'}(t) \rho_{\bq'}(t) \rangle$ 
is a four-point correlator with projected dynamics. On an {\em ad hoc} 
basis, static and dynamic correlations are then subjected to Gaussian 
factorization. An additional convolution approximation for the statics 
reduces the memory function to $\gamma_{\bk}(t) \approx \int d{\bq} \, w_{\bq,\bk-\bq} 
F_q(t) F_{|\bk -\bq|}(t)$, with a weight-factor $w_{\bp,\bq}$ containing 
only static information. Under these approximations for $\gamma_{\bk}(t)$ 
equation (\ref{equ:Fkt}) is closed in $F_k(t)$ and we have arrived at 
standard MCT. 

Various versions of extended mode coupling theories (eMCT) have been 
put forward to account for activated processes \cite{emct}. These 
theories perturbatively invoke a coupling to current 
modes. But recent simulations of Brownian systems \cite{Szamel}, where 
the momentum current should not effectively couple to slow relaxation, 
have shown that deviations from MCT can be as large as they are in 
Newtonian systems. Also, recent theoretical work has cast grave doubt 
on the role of current modes in eMCT \cite{CatGiu}. It would appear 
that perturbative coupling to currents cannot provide the local physics 
necessary to restore ergodicity deep within the activated regime.

Within a theory entirely based on density modes one is thus lead to reconsider 
the applicability of Gaussian factorization \cite{zacc,Szamel2003b,Wu2005}. 
Proceeding in the projection 
operator approach one shows that the actual evolution of four-point correlations 
is governed by an equation analogous to (\ref{equ:Fkt}) that, in turn, couples 
to six-point correlations, and so on. The factorization approximation may thus 
be delayed, a perturbative framework we refer to as generalized mode 
coupling theory (gMCT). Recent implementations of gMCT \cite{Szamel2003b,Wu2005} 
have demonstrated that the inclusion of higher order density correlations appears 
to systematically lower the transition temperature (or raise the transition volume 
fraction), suggesting that some aspect of activated behavior is captured. 

In this letter we illustrate the potential of gMCT to account for 
activated processes, meaning -- here and throughout -- the processes discarded 
in MCT due to factorization. This may be achieved through the {\em non-perturbative} 
limit where the full multi-point basis of dynamical correlations is
included \cite{Szamel2004e}.  
Our discussion is based on a schematic model as its analytic tractability 
provides deeper insights than could be obtained from numerical analysis of 
more realistic systems. To motivate its form we briefly return to 
Eq.~(\ref{equ:Fkt}) and its higher order generalizations. It is useful to 
normalize $\phi_k(t) = F_k(t)/S_k$. For simplicity and following \cite{Szamel2003b,Wu2005} 
we focus on diagonal contributions $\bq = \bq'$ in the memory function $\gamma_{\bk}(t)$. 
Treating static correlations as in MCT but retaining the dynamic four-point 
correlations $\phi_{\bk_1,\bk_2}(t) = 
F_{\bk_1,\bk_2,\bk_1,\bk_2}(t)/F_{\bk_1,\bk_2,\bk_1,\bk_2}(0)$ leads to 
$\gamma_{\bk}(t) \approx \int d{\bq} \, \Lambda_{\bq,\bk-\bq} \phi_{\bq,\bk-\bq}(t)$ 
where $\Lambda_{\bp,\bq} = S_p S_q w_{\bp,\bq}$ with the same weight-factor 
$w_{\bp,\bq}$ as in MCT. At second-lowest order the projection operators 
yield \cite{Szamel2003b,Wu2005}
\begin{equation*}
  \ddot{\phi}_{\bk_1,\bk_2}(t) + \mu_{k_1,k_2} \phi_{\bk_1,\bk_2}(t) 
  + (\gamma_{\bk_1,\bk_2} * \dot{\phi}_{\bk_1,\bk_2})(t) = 0,
\end{equation*}
with $\mu_{k_1,k_2} \approx \mu_{k_1} + \mu_{k_2}$. The fluctuating force 
for pair densities is essentially a product of three density modes. We 
perform a corresponding projection and again only consider diagonal wave vector 
terms in $\gamma_{\bk_1,\bk_2}(t)$. Simple treatment of the static projections as 
before then gives 
$\gamma_{\bk_1,\bk_2}(t) \approx [ 
\mu_{k_1} \int d{\bq} \Lambda_{\bq,\bk_1-\bq} \phi_{\bq,\bk_1-\bq,\bk_2}(t) + 
\mu_{k_2} \times \int d{\bq} \Lambda_{\bq,\bk_2-\bq} \phi_{\bq,\bk_1,\bk_2-\bq}(t)
]/(\mu_{k_1}+\mu_{k_2})$ 
{and $\phi_{\bk_1,\bk_2,\bk_3}(t)$} a normalized six-point correlation function. 

This procedure may be continued indefinitely and induces a hierarchy of evolution 
equations for dynamical multi-point correlations. Our schematic model is obtained 
by dropping the wave-vector indices. We replace $\phi_k(t) \mapsto \phi_1(t)$, 
$\phi_{\bk_1,\bk_2} \mapsto \phi_2(t)$, etc., and since we effectively do not 
discriminate between different wave-vectors $\mu_{k_1} + \ldots + \mu_{k_n} 
\mapsto \mu n$. The memory-functions naturally \cite{footnote} 
become $\gamma_{\bk_1,\ldots,\bk_n}(t) \mapsto \mu \Lambda \phi_{n+1}(t)$. 
Neglecting inertial effects and including the bare viscosities 
arising from the short-time portion of the memory functions 
we arrive at the schematic hierarchy 
\begin{equation}
  \dot{\phi}_n(t)+ \mu n \phi_n(t) + \mu \Lambda (\phi_{n+1} * \dot{\phi}_n)(t)=0. 
  \label{equ:hdef} 
\end{equation}
From now on we set $\mu = 1$ as it can be absorbed by a rescaling of time. 
Note that Eq.~(\ref{equ:hdef}) reduces to the $F_2$ model when closed with 
$\phi_2(t) = \phi_1^2(t)$ \cite{Gotze,leutheusser1984}. 
The drastic simplifications inherent in our schematic hierarchy clearly 
prohibit a quantitative interpretation of its predictions. 
Nevertheless, one would expect that Eq.~(\ref{equ:hdef}) captures the 
general phenomenology of gMCT in a similar way as the $F_2$ model does for MCT. 

The entire analysis of our schematic hierarchy is based on one central identity. 
It applies in Laplace transformed representation, $\phih_n(s) = \mathcal{L}
\{\phi_n(t)\}$, where Eq.~(\ref{equ:hdef}) reads 
\begin{equation}
  \phih_n(s) = \left[ s + \frac{n}{1 + \Lambda \phih_{n+1}(s)} \right]^{-1}. 
  \label{equ:hLap}
\end{equation}
Recursion of this equation produces a continued fraction. Using various 
transformations (all mathematical details will be given in \cite{MayMiyRei06}), 
however, we derive an identity that directly relates any two functions 
$\phih_m$ and $\phih_n$: 
\begin{equation}
	\Omega_m(s) = \Omega_n(s) \quad \mbox{with} \quad \Omega_n(s) = \Gamma(n) 
  \frac{v_n(s)}{u_n(s)}, 
	\label{equ:Omega} 
\end{equation}
where $\Gamma(z)$ denotes the Gamma function and 
\begin{eqnarray*}
  u_n(s) & = & \Psi(s+1,s+n+1;\Lambda) + \phih_n(s) \Psi(s,s+n;\Lambda), \\
  v_n(s) & = & \Phi(s+1,s+n+1;\Lambda) - \phih_n(s) \Phi(s,s+n;\Lambda). 
\end{eqnarray*}
The latter expressions contain the (regularized) confluent hypergeometric 
functions $\Phi(a,b;z) = {_1}F_1(a,b;z)/\Gamma(b)$ and $\Psi(a,b;z)$ \cite{math}. 
Note that $\Omega_m(s) = \Omega_n(s)$ for any $m,n$ implies 
that $\Omega_n(s)$ is independent of $n$ and thus invariant under 
(\ref{equ:hLap}). So any single $\phih_m(s)$ determines the invariant 
$\Omega(s) \equiv \Omega_m(s)$ and from that, in turn, all $\phih_n(s)$ 
follow by simply rearranging $\Omega(s)=\Omega_n(s)$. 

In gMCT the hierarchy (\ref{equ:hdef}) is closed through factorization 
at some order $N \geq 2$. In general this amounts to $\phi_N(t) = 
\phi_{n_1}^{m_1}(t) \phi_{n_{2}}^{m_{2}}(t) \cdots \phi_{n_{k}}^{m_{k}}(t)$. 
Here orders $n_i$ appear with multiplicity $m_i \geq 1$ such that $N = n_1 m_1 
+ n_2 m_2+ \ldots + n_k m_k$. In schematic MCT $\phi_2 = \phi_1^2$ 
while gMCT closures considered previously correspond 
to $\phi_3 = \phi_1 \phi_2$ \cite{Szamel2003b} and $\phi_4 = \phi_1^2 \phi_2$ \cite{Wu2005}. It turns out that 
for {\em any} factorization there is a critical coupling $\Lambda_c$ 
above which the $\phi_n(t)$ do not relax fully, i.e., $q_n = \phi_n(t\to\infty) 
> 0$. This can be deduced from Eq.~(\ref{equ:Omega}): associated 
with each closure is a polynomial, say in $q_1$, whose real roots are dynamical 
fixed points. Some examples of critical couplings $\Lambda_c$, where real roots 
appear, and the corresponding plateau heights $q_1$ for low order closures are 
listed in Tab.~\ref{tab:lc}. 
\begin{table}[!htb] 
  \begin{tabular}{c||c||c|c||c|c|c|c}
                   & $\phi_1^2$        & $\phi_1^3$ & $\phi_1 \phi_2$ 
      & $\phi_1^4$ & $\phi_1^2 \phi_2$ & $\phi_2^2$ & $\phi_1 \phi_3$ \\ 
    \hline 
    $\Lambda_c$    & 4                 & 4.4922     & 4.8284          
      & 4.9398     & 5.2359            & 5.6946     & 5.6046          \\
    $q_1$          & $\frac{1}{2}$     & 0.5453     & 0.5858          
      & 0.5839     & 0.6205            & 0.6782     & 0.6565 
  \end{tabular} 
  \caption{\label{tab:lc} Critical couplings $\Lambda_c$ and associated 
    plateaus $q_1 = \phi_1(t \to \infty)$ for all factorizations of 
    $\phi_N(t)$ with $N=2,3,4$.} 
\end{table} 
We observe that within each order $N$ the value of $\Lambda_c$ is lowest 
for the factorization $\phi_N=\phi_1^N$ while it is largest for 
$\phi_N = \phi_{n} \phi_{N-n}$ with $n$ the integer part of $N/2$. 
More importantly there is an 
overall increase in $\Lambda_c$ as we raise $N$. Indeed, for 
the factorization $\phi_N=\phi_1^N$ it can be shown that $\Lambda_c \sim 
N$ for $N \gg 1$. Including higher order correlations does not merely 
affect $\Lambda_c$ quantitatively but in fact allows us to increase its 
value {\em unboundedly}. Implications of this result will become clear 
in the following. 

The dynamics of gMCT under higher order factorization closures, 
which has never been investigated, can be obtained (schematically) 
by numerical integration of Eq.~(\ref{equ:hdef}). Due to factorization of 
$\phi_N(t)$ the hierarchy reduces to a set of coupled non-linear 
integro-differential equations in $\{\phi_1(t), \ldots, 
\phi_{N-1}(t)\}$. It is efficiently integrated by the algorithm 
of \cite{FucGotHofLat91}. The results in gMCT with $N \geq 3$ 
turn out to be remarkably similar to those of 
MCT. For any given factorization, $\phi_1(t)$ displays two-step 
relaxation when $\Lambda$ approaches the relevant critical value 
$\Lambda_c$ from below. As $\Lambda \to \Lambda_c$ the 
$\alpha$-relaxation time $\tau$ diverges and $\phi_1(t)$ assumes 
the non-zero limit $q_1 = \phi_1(t \to \infty)$. We have 
performed a scaling analysis, which is based on Eq.~(\ref{equ:Omega}) and 
$\phih_N(s) = \mathcal{L}\left\{ \prod_i \phi_{n_i}^{m_i}(t) \right\}$,  
to precisely characterize the critical properties near $\Lambda_c$. 
It can be shown that the scaling equations for 
our hierarchy are in fact formally identical to those for usual schematic 
models \cite{Gotze}. Hence $\beta$-relaxation follows the power-laws 
$\phi_1(t) - q_1 \sim t^{-a}$ and $q_1 - \phi_1(t) \sim t^b$ in the 
early and late regimes, respectively, and the $\alpha$-relaxation time 
diverges like $\tau \sim 1/(\Lambda_c - \Lambda)^{\gamma}$. Strikingly 
the exponents $a,b$ and $\gamma$ for these scaling laws are {\em identical} 
for {\em all} factorization closures $\phi_N = \prod_i \phi_{n_i}^{m_i}$. 
Their values exactly match those of the $F_2$ model \cite{Gotze}, that 
is $a \approx 0.395$, $b = 1$ and $\gamma = \frac{1}{2a} + \frac{1}{2b}$, 
regardless of closure. 

\begin{figure}
  \epsfig{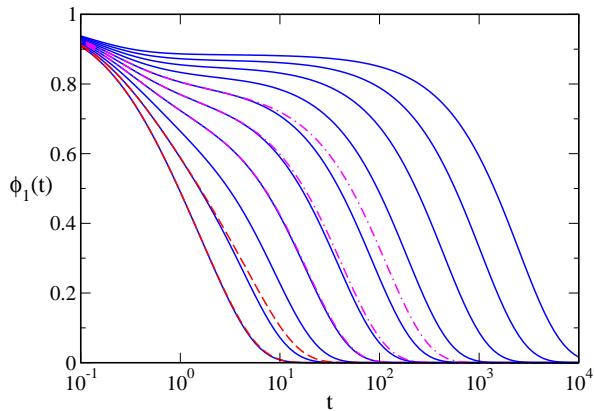} 
  \caption{\label{fig:phi_t} [color online] Exact solutions 
    of the infinite hierarchy, Eq.~(\ref{equ:phiinf}), at $\Lambda = 
    1,2,\ldots, 10$ (full curves), MCT with closure 
    $\phi_2=\phi_1^2$ at $\Lambda = 1,2$ (dashed) and gMCT 
    under $\phi_7=\phi_3 \phi_4$ at $\Lambda = 4,5,6$ (dashed-dotted). 
    See text for discussion.}
\end{figure}

While tempting, the above results cannot be interpreted as supporting 
the standard MCT picture. The occurrence of the robust critical scaling 
under higher order closures is systematically deferred along with the 
critical point $\Lambda_c \sim N$ itself. 
We turn to the non-perturbative limit $N \to \infty$ of avoided 
factorization to develop a deeper understanding. According 
to our general result (\ref{equ:Omega}) all functions $\phih_n(s)$ 
are known once we have determined the invariant $\Omega(s)$. 
{\em A priori} the latter depends on how we close 
the hierarchy since we may write $\Omega(s) = \Omega_N(s)$. 
However, analysis of $\Omega_N(s)$ shows that for all physically 
reasonable closures  
the invariant of the infinite hierarchy vanishes $\Omega(s) 
= \lim_{N \to \infty} \Omega_N(s) = 0$. Thus the infinite hierarchy has a 
{\em unique} solution. Rearranging $\Omega_n(s) = \Omega(s) = 0$ 
then produces the exact result 
\begin{equation} 
  \phih_n(s) = \frac{\Phi(s+1,s+n+1;\Lambda)}{\Phi(s,s+n;\Lambda)}. 
  \label{equ:phiinf} 
\end{equation} 
The regularized confluent hypergeometric functions $\Phi$ are 
analytic in $s$. Therefore the only singularities in Eq.~(\ref{equ:phiinf}) 
are (first order) poles at points $\{s_i\}$ where the denominator 
$\Phi(s_i,s_i+n;\Lambda)=0$ vanishes. The relaxation spectrum 
$\{s_i\}$ is infinite, discrete and contained in the negative 
real axis. Consequently the inverse Laplace transform of Eq.~(\ref{equ:phiinf}) 
is of the form $\phi_n(t) = \sum_{i=0}^\infty r_i e^{s_i t}$ with 
$r_i$ the residues at $s = s_i$. Numerical evaluation of both 
$\{s_i\}$ and $\{r_i\}$ is straightforward. Plots of the solutions 
$\phi_1(t)$ of the infinite hierarchy are shown in Fig.~\ref{fig:phi_t}. 
Upon increasing $\Lambda$ a single component $s_0$ of the spectrum 
approaches the origin. Hence the shape of $\alpha$-relaxation  in our 
schematic hierarchy is exponential 
with $\tau$ defined by $s_0 t = -t/\tau$. From $\Phi(s_0,s_0+1;\Lambda)=0$ 
one shows that $\tau \sim e^\Lambda/\Lambda$ for $\Lambda \gg 1$. 
There is {\em no} MCT {\em transition} in the infinite hierarchy at any 
finite $\Lambda$. Instead, the relaxation time $\tau$ essentially grows 
{\em exponentially} in $\Lambda$. Remarkably this hallmark of non-perturbative 
behavior \cite{instant} emerges from our microscopically 
motivated dynamical approach. 

We now discuss how the above results merge into a consistent 
picture. In the infinite hierarchy, memory effects first appear in the 
relaxation of $\phi_1(t)$ but gradually spread into the hierarchy 
upon increasing $\Lambda$. This causes the exponential slowdown in the 
dynamics of $\phi_1(t)$. The MCT closure $\phi_2 = \phi_1^2$, while 
appropriate for small $\Lambda$, generally overestimates the relaxation 
time of $\phi_2(t)$. As shown 
in Fig.~\ref{fig:phi_t} the MCT solution matches that of the infinite hierarchy 
at $\Lambda = 1$ but already deviates at $\Lambda = 2$. When 
$\Lambda$ approaches $\Lambda_c = 4$ the feedback through the closure drives 
the MCT solution into a spurious transition while, in the infinite hierarchy, 
the relaxation of 
$\phi_1(t)$ only develops a shoulder (see Fig.~\ref{fig:phi_t} at $\Lambda=4$). 
This is also typically observed in 
experiments and simulations at the point where MCT {\em predicts} a 
glass transition \cite{Kob}. If, however, the factorization is delayed to higher 
order correlations, then so is the appearance of 
feedback through the closure. We refer again to Fig.~\ref{fig:phi_t}, which 
also shows gMCT solutions under factorization $\phi_7 = \phi_3 \phi_4$. 
These perfectly match the infinite hierarchy even at $\Lambda = 4$, but 
again deviate as $\Lambda$ is raised further towards the critical value 
$\Lambda_c \approx 8.1049$ of this closure. 
Of course the validity of gMCT may be extended to any value $\Lambda$ 
by raising $N$. In the non-perturbative limit 
the transition is deferred to $\Lambda_c \sim N \to \infty$ and we arrive 
at the ergodic solution (\ref{equ:phiinf}) of the infinite hierarchy. 
The fact that the $\phi_n(t)$ derived from Eq.~(\ref{equ:phiinf}) do 
not obey any factorization corresponds to underlying {\em non-Gaussian} 
dynamical fluctuations \cite{Szamel2004e}. In microscopic systems these
are a natural consequence of cooperative, heterogeneous dynamics -- 
not fully accounted for in MCT due to Gaussian factorization. However, microscopic 
gMCT offers the potential to systematically capture non-Gaussian fluctuations 
and thus the underlying processes that cause them. 

\begin{figure}
  \epsfig{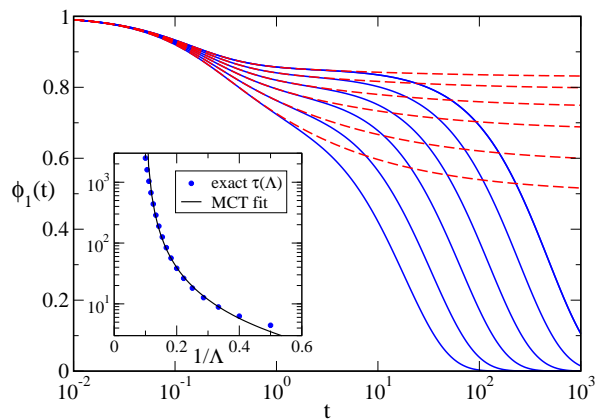} 
  \caption{\label{fig:mct} [color online] Main Panel: critical solutions 
  (dashed) for closures $\phi_1^2$, $\phi_1 \phi_2$, $\phi_2^2$, $\phi_2 \phi_3$, 
  $\phi_3^2$, $\phi_3 \phi_4$ at $\Lambda_c \approx 4$, 4.8, 5.7, 6.5, 7.3, 8.1, 
  respectively, and solutions of the infinite hierarchy at the same $\Lambda$ (full 
  lines). Inset: $\alpha$-relaxation time of the infinite hierarchy (dots) for 
  $\Lambda = 2$, 2.5, \ldots, 10 and an MCT fit (full curve), 
  $\tau_\mathrm{fit} = A/(\Lambda^{-1}-\Lambda_c^{-1})^{\gamma}$, with $\gamma 
  \approx 1.765$ fixed but $A \approx 0.7$ and $\Lambda_c \approx 10.1$ fitted.}
\end{figure}

Strikingly the solutions of the infinite hierarchy nevertheless carry some important 
features consistent with MCT. Figure~\ref{fig:mct} shows gMCT solutions under 
factorization at orders $N=2,3,\ldots, 7$ at the corresponding critical couplings 
$\Lambda_c$. From our discussion of the dynamics in gMCT we know that each of 
these functions approaches its relevant plateau $q_1$, see Tab.~\ref{tab:lc}, 
like a power law $t^{-a}$ with the MCT exponent $a \approx 0.395$. Now compare 
these critical solutions to those of the infinite hierarchy, Fig.~\ref{fig:mct}, 
at the same values of $\Lambda$: apparently there is an excellent match 
throughout the early $\beta$-relaxation regime. In this sense the hierarchy 
{\em appears} to approach criticality with a $\beta$-relaxation exponent $a$ 
as predicted by MCT. Likewise, we could extract the MCT exponent $b=1$ 
characterizing late $\beta$-relaxation where $\phi_1(t)$ drops below its plateau. 
A consistent MCT picture would then require 
power-law scaling of the relaxation time $\tau$ with a particular exponent 
$\gamma$. 
For fitting purposes this power-law is usually expressed 
in $T-T_c$ rather than $\Lambda_c - \Lambda$, which is 
equivalent (asymptotically) close to criticality. 
Noting that $1/\Lambda$ is generally an increasing function of $T$ \cite{Gotze}
(and thus represents a temperature-like variable) we write 
$\tau_\mathrm{fit}=A/(\Lambda^{-1}-\Lambda_c^{-1})^{\gamma}$. 
Most remarkably 
this power law with the MCT exponent $\gamma \approx 1.765$ 
indeed fits the relaxation time of the infinite hierarchy well over 
$3 \leq \Lambda \leq 9$ covering, as in simulations \cite{Kob}, 
about two to three decades in $\tau$, see inset Fig.~\ref{fig:mct}. 
Thus, within this limited dynamical range the solutions of the 
infinite hierarchy exhibit power-law behaviors characterized by the 
MCT exponents $a,b$ and $\gamma$. 
The fitted $\Lambda_c \approx 10.1$, lying 
well above the MCT value $\Lambda_c = 4$, is consistent with correspondingly 
lower $T_c$ extrapolated in experiments and simulations \cite{Kob}. 
However, contrary to MCT these power-laws do not reflect asymptotic 
scaling in the infinite hierarchy. In fact, there is no divergence of 
the relaxation time $\tau \sim e^\Lambda/\Lambda$ at any finite $\Lambda$, 
and in particular not at $\Lambda_c \approx 10.1$ as extrapolated from the 
power-law fit. 

The entire phenomenology discussed above emerges purely from the structure of 
our schematic hierarchy (\ref{equ:hdef}). The qualitative differences between 
the non-perturbative and MCT solutions of Eq.~(\ref{equ:hdef}) in several aspects 
resemble those found when contrasting experimental or simulation data 
\cite{Kob,expsim,Szamel} with microscopic  ($\bk$-dependent) MCT, e.g., predicted 
versus fitted $T_c$ or the apparent power-law behaviors and deviations away 
from them. Our analysis resolves these discrepancies, leading to a 
coherent picture based upon non-Gaussian dynamical fluctuations. 
To describe the relaxation of supercooled liquids in full detail, including, 
e.g., stretched exponentials, numerical analysis of microscopic ($\bk$-dependent) 
gMCT will be necessary. One would hope that such an approach preserves the 
successful features of MCT, for instance, accurate prediction of the plateau values. 
Beyond that, however, it might not only offer a possibility 
to systematically improve the predictions of MCT, particularly through the non-perturbative limit, but 
also provide a kinetic approach to capture dynamical heterogeneity. 
The implications of the work presented here on related issues, such as the breakdown 
of the Stokes-Einstein relation, will be explored in future work. 

We thank H.\ C.\ Andersen, G.\ Biroli, J.\ P.\ Bouchaud, M.\ E.\ Cates, W.\ G\"{o}tze, 
and G.\ Szamel for discussions, and acknowledge support from the NSF \# 0134969.


\begin{thebibliography}{10}
\bibitem{footnote}
We replace $\int d\bq \Lambda_{\bq,\bk-\bq}$ by an effective weight $\mu \Lambda$ 
(and similarly assume uniformity of the short-time viscosities).
\end{thebibliography}
\end{document}